\documentclass[12pt,preprint]{aastex}
\usepackage{epsfig}
\usepackage{graphicx}
\usepackage{longtable}
\usepackage{rotating}
\usepackage{palatino,  multicol}
\usepackage{natbib}
\usepackage{latexsym}
\bibpunct{(}{)}{;}{a}{}{,}

\title{The Island state of the Atoll Source  4U 1820--30}

\author{Diego Altamirano\altaffilmark{1},  M. van der Klis\altaffilmark{1},\\  M. M\'endez\altaffilmark{2}, 
S. Migliari\altaffilmark{1}, 
P.G. Jonker\altaffilmark{2,3}, 
A. Tiengo\altaffilmark{4,5} 
W. Zhang\altaffilmark{6}.} 

\email{diego@science.uva.nl}

\altaffiltext{1}{Astronomical Institute, ``Anton Pannekoek'', University of Amsterdam, and Center for High Energy Astrophysics, Kruislaan 403, 1098 SJ Amsterdam, The Netherlands.}

\altaffiltext{2}{SRON, National Institute for Space Research, Sorbonnelaan 2, 3584 CA, Utrecht, The Netherlands.}

\altaffiltext{3}{Harvard-Smithsonian Center for Astrophysics, 60 Garden Street, MS 83, MA 02138, Cambridge, U.S.A.}

\altaffiltext{4}{Istituto di Astrofisica Spaziale e Fisica Cosmica, Sezione di Milano ``G.Occhialini'', via Bassini 15, I-20133 Milano, Italy.}

\altaffiltext{5}{Universit\`a degli Studi di Milano, Dipartimento di Fisica, via Celoria 16, I-20133 Milano, Italy.}

\altaffiltext{6}{Laboratory for X-Ray Astrophysics, NASA Goddard Space Flight Center, Greenbelt, MD 20771, U.S.A.} 

\date{}

\begin{document}

\begin{abstract}

We study the rapid X-ray time variability in all public data available from 
the \textit{Rossi X-ray Timing Explorer's} Proportional Counter Array on
the atoll source  4U 1820--30 in the low-luminosity island state.
A total of $\sim46$ ks of data were used. 
We compare the frequencies
of the variability components of  4U 1820--30 with those in other atolls sources.
These frequencies were previously found to  follow a universal scheme of correlations. 
We find that 4U 1820--30 shows   
correlations that are shifted by factors of $1.13\pm0.01$ and $1.21\pm0.02$ with
respect to those in other atoll sources. These shifts are similar to, but smaller than the shift factor
$\sim1.45$ previously reported for some accreting millisecond 
pulsars. 
Therefore, 4U 1820--30 is the first atoll source which
shows no significant pulsations but has a significant shift in the frequency correlations
compared with other 3 non-pulsating atoll sources.
\end{abstract}

\maketitle

\section{Introduction}
\label{sec:intro}

Accretion in neutron star low-mass X-ray binaries (LMXBs) can be
studied through the spectral and timing properties of the associated X-ray emission. 
The Fourier power spectra of the X-ray flux of 
these systems, exhibit quasi-periodic oscillations (QPOs) as well as noise components
between $\sim1\times10^{-3}$ Hz and
$\sim1350$ Hz. Most of these variability components are thought to be associated with processes
in the accretion disk \citep[for reviews and references see][]{Vanderklis00,Vanderklis04}
but some of them may arise  on the surface 
of the neutron star \citep[see e.g.][]{Revnivtsev01,Strohmayer03}.
 The timing properties at low frequencies ($\nu < 100$ Hz) as well as the spectral properties are the basis of 
the classification of these systems as either Z or atoll sources \citep{Hasinger89}. 
In recent literature, each variability component is designated
 $L_i$ -- L for 'Lorentzian', \citet{Belloni02} -- where the index i indicates
 the component; the component's characteristic frequency is designated $\nu_i$. For example, $L_b$ 
is an often flat-topped broad band noise component at a low frequency $\nu_b$, and 
$L_u$ the upper kilohertz QPO with frequency $\nu_u$ \citep[see][for complete terminology]{Straaten03}.

 The kilohertz QPOs are seen between 
a few hundred and $\sim1350$ Hz and when two of them are seen at the same time (twin kHz QPOs), 
the difference between their frequencies is constrained between $\sim185$ Hz and $\sim400$ Hz. 
In the 0.01-100 Hz range two to five band-limited noise, peaked-noise and QPO components
 are observed whose frequencies 
all correlate with one another and with that of the kilohertz QPOs 
\citep[see][ and references within]{Straaten05}.
An example is the WK correlation \citep[after][]{Wijnands99}, 
between the hump frequency
 $\nu_h$ and
the break  frequency $\nu_b$. 
This relation may be fundamental in the  understanding of the processes of accretion in  
LMXBs  because, in  
 atoll sources and black holes $\nu_b$ and $\nu_h$ correlate over 3 orders of magnitude 
(the Z sources have slightly higher $\nu_h$).
The existence of such correlations suggests that similar physical phenomena
may be responsible for some of the QPOs and noise components found over
wide ranges of frequency and coherence in Z, atoll and black hole sources.

\citet{Straaten05} found that the frequencies of the noise and QPO components of the accreting millisecond 
pulsar SAX J1808.4--3658 also correlate with $\nu_u$, but, in a different way than  
 those of the other atoll sources. 

They interpreted the difference between the pulsar and the atoll sources
as due to a shift in frequency of the upper kilohertz QPO and  
suggested that physical differences between these sources are most likely to affect the high
frequency components.
In SAX J1808.4-3658, the factors by which $\nu_u$ had to be multiplied to make the correlations 
coincide with those of the ordinary atoll sources were $1.420\pm0.013$ for $\nu_b$, and 
$1.481\pm0.013$ for $\nu_h$.

4U 1820--30 is a low-mass X-ray binary with an orbital period of only 11.4 minutes 
\citep{Stella_PW87} and an X-ray burst source 
\citep[][]{Grindlay76}.  
It is located in the globular cluster NGC\,6624 at a distance $7.6\pm0.4$ kpc \citep{Kuulkers03}.
Radio emission has also been detected from the source \citep{Geldzahler83,Migliari04}.
4U 1820--30 undergoes a regular $\sim$ 176 day accretion cycle \citep{Priedhorsky_T84}, 
switching between high and low states differing by a factor 
$\sim$3 in luminosity \citep{Strohmayer_B02}.
The ultra-compact nature of the system requires that the secondary is a low-mass helium dwarf \citep[see e.g.][]{Rappaport_JN87} so
 that the accreted material likely has a very low hydrogen abundance. 
\citet{Hasinger89} defined 4U 1820--30 as an atoll source and eight years later 
\citet{Smale_ZW97} reported the discovery of kHz quasi-periodic oscillations. 
\cite{Zhang_SSS98} reported the result from a long-term monitoring data set obtained with the Rossi X-ray Timing Explorer.  
They observed kHz QPOs
in both the lower banana and the island state \cite[see][for nomenclature]{Vanderklis04}. They showed that the frequency of the kilohertz QPOs is 
correlated with the PCA count rate below a critical value ($\sim 2500$ counts s$^{-1}$ per 5 PCUs). 
Above this, the QPO frequencies remained constant while 
the count rate increased between $\sim 2500$  and $\sim3200$ counts s$^{-1}$ per 5 PCUs.
{Saturation of QPO frequency at high mass accretion
rates is an expected signature of the marginally stable orbit \citep{Miller98,Kaaret99}; however, 
this is the only source reported to have  shown this behavior, and 
to what extent count rate is  
a good indicator of accretion rate remains to be 
seen \citep[see, e.g. the discussion of the issue in][]{Vanderklis01}.
Later, similar  analysis were carried out, using instead of the count rate, (i) the energy flux, (ii) the 
X-ray spectral shape  \citep{Kaaret99} and 
(iii) the parameter $S_a$  \citep{Bloser_GKZSB00} which  
parameterizes atoll source location in the track of the color-color diagram.  
The same behavior as that observed by \citet{Zhang_SSS98} as a function of count rate was found, 
when the QPO frequency was plotted as a function of either of these three parameters.
The saturation of QPO frequency was interpreted as strong additional evidence for the detection of the marginally
stable orbit in the accretion disk of 4U 1820--30. However, since then \citet{Mendez02} has argued
 that the evidence of the saturation is not so compelling,
especially when some instrumental corrections are taken into account. 
A general tendency
for QPO frequency to saturate toward higher luminosity may be a feature of the same phenomenon
that produces the parallel tracks in frequency-luminosity diagrams \citep{Vanderklis01}.

In this paper, we report on the eight observations that are currently available 
of 4U 1820--30 in the island state. 
All previous works mentioned above included only one observation (20075-01-05-00) of the source in the island state, 
so the current analysis better 
allows us to constrain the power spectral components in the island state of 4U 1820--30 more accurately.
We study the correlations between the characteristic frequencies
of the various timing features, and compare these with those of four well-studied atoll sources, 
three low-luminosity bursters,
one Z-source and  one accreting millisecond pulsar. 
We show that the correlations between frequencies in 4U 1820--30 are shifted 
as found for SAX J1808.4--3658, but with a lower shift factor.
We finally discuss whether the interpretation of a  multiplicative shift of frequencies is the right explanation for the differences
in frequency behavior between the millisecond accreting pulsar SAX J1808.4--3658 and the ordinary atoll sources.

\section{Observations and data analysis}
\label{sec:data}
We used all public data available from the Rossi X-ray Timing Explorer's (RXTE) Proportional Counter Array \citep[PCA; for instrument 
information see][]{Zhang93}. There were 158 pointed observations in 9 programs (10074, 10075, 10076, 20075, 30053,
30057, 40017, 40019, 60030, 70030 and 70031). In our analysis, we use the 16-s time-resolution Standard 2 mode data to calculate X-ray colors. 
For each of the five PCA detectors (PCUs) we calculate a hard and a soft color defined as the count rate in the 9.7--16.0 keV band divided 
by the rate in the 6.0--9.7 keV band and the 3.5--6.0 keV rate divided by the 2.0--3.5 keV rate, respectively. 
For each detector we also calculate the intensity,
defined as the count rate in the energy band 2.0--16 keV. 
To obtain the count rates in these exact energy ranges, we interpolate linearly 
between count rates in the PCU channels.We then subtract the background
 contribution in each band using the standard bright source background model for the PCA, version 2.1e
\footnote{PCA Digest at http://heasarc.gsfc.nasa.gov/ for details of the model}. 
No deadtime corrections were made as the effect of deadtime can be neglected for our purposes ($<0.001\%$).
We calculate the colors and intensity for each time interval of 16s. 
In order to correct for the gain changes
as well as the differences in effective area between the PCUs themselves, we used the method introduced by \citet{Kuulkers94}: for each PCU we 
calculate, in the same manner as for 4U 1820--30, the colors of the Crab which can be supposed to be constant. We then average 
the 16s Crab colors 
and intensity for each PCU for each day. For each PCU we divide the 16s color and intensity values obtained for 4U 1820--30 by the corresponding 
Crab values that are
 closest in time but in the same RXTE gain epoch.
The RXTE gain epoch changes with each new high voltage setting of the PCUs \citep{Jahoda96}.
 After the Crab normalization is done, we average the colors and intensity over all PCUs.
 Finally, we average the
16s colors per observation. 
Figure \ref{fig:ccd} shows the color-color diagram
of the 158 different observations that we used for this analysis, and Figure \ref{fig:cvsint} the
corresponding hardness-intensity diagrams (soft and hard color vs. intensity).

We find 8 observations which are
located in the island region of the color-color diagram (hard colors greater than 0.9).
These observations are the
subject of this paper (see Table \ref{table:obs}).

\begin{table}[!hbtp]
\center
\begin{tabular}{ccccccc}\hline 
Hard & Observation  & Label & Date of & Duration & Total Average  & Average count rate\\ 
Color& ID && Observation & (ks) &count rate (c/s)&  per PCU (c/s/PCU)\\
\hline
1.018 & 40017-01-24-00 & A & Jun-04-2003 & $\sim8.2$  & $\sim720$ & $\sim268$ \\
1.010 & 70030-03-04-00 & B & Jun-11-2003 & $\sim3.2$ & $\sim1125$ & $\sim281$ \\
1.014 & 70030-03-05-00 & B & Jun-14-2003 & $\sim6.5$ & $\sim925$  & $\sim308$ \\
0.993 & 70030-03-04-01 & C & Jun-12-2003 & $\sim6.5$ & $\sim988$  & $\sim283$ \\
1.010 & 70030-03-05-01 & C & Jun-15-2003 & $\sim6.6$ & $\sim1076$ & $\sim358$ \\
0.982 & 70031-05-01-00 & D & Jun-14-2002 & $\sim3.1$  & $\sim1486$& $\sim297$ \\
0.922 & 20075-01-05-00 & E & May-02-1997 & $\sim8.5$  & $\sim1770$& $\sim354$ \\
0.946 & 70030-03-05-02 & E & Jun-16-2003 & $\sim3.2$ & $\sim1260$ & $\sim421$ \\
\hline
\hline
\end{tabular}
\caption{The eight observations used for the timing analysis. The statistical errors in hard color are $\lesssim 0.001$}.
\label{table:obs}
\end{table}

For the Fourier timing analysis of these 8 observations we used an $125\mu$s time resolution Event mode (E\_125us\_64M\_0\_1s). 
Leahy-normalized  power spectra were constructed using data segments of 128 seconds 
and 1/8192s time bins  such that the lowest available frequency is 
$1/128 \approx 8 \times 10^{-3}$ Hz and the Nyquist frequency 4096 Hz.  
Detector drop-outs were removed but no background or deadtime corrections 
were performed prior to the calculation of the power spectra.
We first averaged the power spectra per observation. 
We  inspected the shape of the average power spectra at high frequency ( $>2000$ Hz)
 for unusual features in addition to the usual Poisson noise. None were found. 
We then  subtracted a Poisson noise spectrum  estimated  from the power between 3000 
and 4000 Hz, where neither intrinsic noise nor QPOs are known to be present, using the method developed by \citet{Kleinwolt04} 
based on the analytical function of \citet{Zhang95}.
The resulting power spectra were then converted to squared fractional 
rms \citep{vanderklis95b}. 
In this normalization the square root of the integrated power density is a direct measurement of the variance caused by the
intrinsic 
variability in the source count rate. In three cases it was possible to add two observations together to improve statistics. This 
was done only for those observations
which had similar colors and power spectra consistent with being the same within errors. The resulting power spectra are labeled 
from A to E (Figure \ref{fig:powerspectra}) in order of decreasing hard color. Table \ref{table:obs} shows 
which observations were used to create each of the averaged power spectra. 

To fit the power spectra, we used a multi-Lorentzian function: the sum of several Lorentzian components
plus, if necessary, a power law to fit the very low frequency noise (VLFN). 
Each of these components, is usually described with an $L_i$ ( for 'Lorentzian' ) and its frequency, with $\nu_i$, 
where $i$ determines the type of component. For example, $L_u$ identifies the upper kHz QPO and $\nu_u$ its frequency.
By analogy, other components go by names such as  $L_{\ell}$ (lower kHz), $L_{hHz}$ (hectohertz),
$L_h$ (hump), $L_b$ (break frequency), and their frequencies as $\nu_{\ell}$, $\nu_{hHz}$, $\nu_h$ and $\nu_b$, respectively.
Using this multi-Lorentzian function makes it straightforward to directly  compare the different components in 4U 1820--30
to those in previous works which used the same fit function
\citep[e.g.,][and references therein]{Belloni02,Straaten03,Straaten05}.

In the fits we only include those Lorentzians with a significance larger than $3\sigma$
 based on the error in the power integrated from 0 to $\infty$.
We give the results of the fits in terms of $\nu_{max}$ and $Q$, of which 
$\nu_{max}$ was introduced by \citet{Belloni02} as $\nu_{max} = \sqrt{\nu^2_0 + (\frac{FWHM}{2})^2} = \nu_0 \sqrt{1 + \frac{1}{4Q^2}}$. 
For Q we use the standard definition $Q = \frac{\nu_0}{FWHM}$. FWHM is the full width at half
maximum of the Lorentzian.

\section{Results}\label{sec:results}
Figures \ref{fig:ccd} and \ref{fig:cvsint} show that in order A to E, the spectrum becomes 
softer, i.e. both hard and soft color decrease, 
 while the spectrum is harder than in the banana branch and the 
intensity is approximately constant (see Figure \ref{fig:cvsint}).
This is the expected behavior for an atoll source which
is moving from
the island to the lower left banana state in the color color diagram \citep{Vanderklis04}.

In Figure \ref{fig:powerspectra}, we show the average power spectra with their fits. 
Four to five Lorentzian 
components were needed for a good fit
of power spectra \textbf{A--D }.
Power spectrum \textbf{E}, whose colors are closest to the upper banana state, 
could be fitted with either six or seven Lorentzians. Both fits share six components whose frequencies  are the same within 
errors; in the case of 7 Lorentzians,  
an extra component is present at $\nu_3 = 407.9\pm30.5$. This component is significant only at
$\sim2\sigma$ (single trial) level, and represents an $\sim1.3\sigma$ 
improvement of the $\chi^2$ of the fit according to an F-test. 
However, if this component, which is consistent with being the lower kilohertz QPO peak, 
is not included in the model, 
the fit becomes unstable unless the quality factor $Q_{hHz}$ is fixed. 

Table \ref{table:data} gives the results of the fits to the power spectra and 
in  Figure \ref{fig:nuvsnu}, we show 
the correlations of the characteristic frequencies $\nu_{max}$ of the power
spectral components with the frequency of the upper kilohertz QPO $\nu_{u}$. 
For power spectrum E, we always show the results for 7 Lorentzians.

As expected for the island state of an atoll source, $\nu_{u}$ is lower than $\sim700$ Hz 
  \citep[see e.g.][]{Straaten03,Straaten05,Vanderklis04} and increases monotonically from A to E with decreasing hard color.
 $L_{hHz}$ is at similar frequencies as in  
the other atoll sources, between $\sim100$ and $\sim200$ Hz.
 
For $L_{b}$ and $L_{h}$,  
a shift appears to exist between the correlations of 4U 1820--30 
and those of the other atoll sources 
studied by \citet{Straaten05}.
To further investigate this, in Figures \ref{fig:b} and \ref{fig:h} we plot  
$\nu_b$ and $\nu_h$ respectively, 
versus $\nu_u$. We use all the data used by \citet{Straaten05}
for the atoll sources and the 
low luminosity bursters; however, of the millisecond pulsars, 
we only use data of
 SAX J1808.4--3658,  
which, in contrast to the others, has data points in the same frequency
region as 4U 1820--30.
As can be seen in Figures \ref{fig:b} and \ref{fig:h}, our points for 4U 1820--30 are right in the important transition 
region around $\nu_u \sim 600$ Hz. On one hand for $L_b$ (Figure \ref{fig:b}),  our points seem to link 
the SAX J1808.4--3658  data with those for the atoll sources with $\nu_u \gtrsim 600$ Hz. 
However, neither 
the frequency range covered by 4U 1820--30 nor SAX J1808.4--3658 is sufficient to draw the conclusion that the 
two different correlations below $\nu_{u} \sim 600$ Hz become the same correlation above $\nu_{u} \sim 600$ Hz,
as the figure seems to suggest. On the other hand, as shown in Figure \ref{fig:h}, in 4U 1820--30 the
$L_h$ points seem to lie between those of the atoll sources and those of SAX J1808.4--3658.

To determine the shift factors between the frequency correlations of 4U 1820--30
and those of the other atoll sources, 
and to be able to compare them with the shift factors found for SAX J1808.4--3658, 
we followed the same procedure as used by \citet{Straaten05}: we considered the $\nu_b$ vs. $\nu_u$ 
and $\nu_h$ vs. $\nu_u$ relations for which $\nu_u < 600$ Hz, as the behavior of the low-frequency components above 600 Hz is complex.
In practice, this means that we exclude power spectrum E. Note that in our analysis we included the data point
for SAX J1808.4--3658 at $\nu_u = 497.6\pm6.9$ Hz that, when shifted, ends up above 600 Hz, and which was excluded by \citet{Straaten05}.

For each relation, 
we fit a power law to the 4U 1820--30 frequencies together with those
of the atoll sources using the FITEXY routine by
\citet{NumericalRecipes}, which performs a straight line fit to data with errors in both coordinates. 
We took the logarithm of the frequencies so that fitting a power law  becomes equivalent to fitting a straight line. 
Before fitting, we multiplied the 4U 1820--30 $\nu_u$ values with a shift factor that 
ran between 0.1 and 3 with steps of 0.001. The fit with the minimal 
$\chi^2$ then gave the best shift factor. The errors in the shift factor use 
$\Delta\chi^2 = 1$, corresponding to a 68\% confidence level.

The best shift factors in $\nu_u$ for 4U 1820--30 are $1.21\pm0.02$ ($\chi^2/dof = 19.4/18$) and $1.13\pm0.01$ ( $\chi^2/dof =45.3/18$) 
for $\nu_b$ and $\nu_h$ respectively. 

If we repeat the procedure described above, 
but this time instead of multiplying $\nu_u$, we multiply $\nu_b$ and $\nu_h$ by a variable factor 
(vertical frequency shifts in Figure \ref{fig:nuvsnu}) 
,
the best shift factor in $\nu_b$ is $0.55\pm0.03$ ($\chi^2/dof = 19.4/18$) and 
in $\nu_h$ $0.73\pm0.02$ ( $\chi^2/dof =45.3/18$). 
Clearly, the high $\chi^2$ when calculating the best fit for $\nu_h$ indicates 
that the dispersion of the data around the power law is larger than
expected from counting statistics.

In Figure \ref{fig:bvsh} we plot the characteristic frequency $\nu_h$  versus $\nu_b$. As \citet{Straaten05} showed,
the millisecond pulsar SAX J1808.4--3658  follows approximately the same correlation as the atoll sources and low luminosity 
bursters at frequencies  $\nu_b \lesssim 3$ Hz. For $3 \lesssim \nu_b \lesssim 5$ Hz, the atoll sources slightly deviate,
 as $\nu_b$ increases, toward lower $\nu_h$. For $\nu_b \gtrsim 5$, \citet{Straaten05} observed a non-continuous bifurcation 
where $\nu_b$ of the atoll sources  jumps to higher frequencies while SAX J1808.4--3658
 smoothly extends the correlation observed for $\nu_b \lesssim 3$ Hz. 
Our new data for 4U 1820--30, which are all at $\nu_b > 5$ Hz, seem to be in between these two correlations, 
apparently following the behavior of the atoll sources for $3 \lesssim \nu_b \lesssim 5$ Hz. However, 
the point for 4U 1820--30 at higher $\nu_b$ (and higher $\nu_h$), falls in the correlation of SAX J1808.4--3658. 

In Figure \ref{fig:bvsh} we also show the frequency of the Horizontal Branch Oscillation (HBO) and its
subharmonic versus that of the Low Frequency Noise (LFN) for the Z-source GX 5--1. 
The data of GX 5--1  was taken from \citet{Straaten03} \citep[but see][ for original data]{Jonker02}.
The HBO component of GX 5--1 follows the same correlation as SAX J1808.4--3658 but, as already noted
by \citet{Wijnands99}, the HBO of Z-sources is slightly higher in this diagram than the $L_h$ and $L_{LF}$
components of atoll sources.
The HBO subharmonic extends the correlation
that is found for atoll sources and low luminosity bursters for $\nu_b \gtrsim 5$ Hz to lower frequencies, 
suggesting that the apparent bifurcation mentioned before could be associated with  harmonic mode switching.

In Figure \ref{fig:bumps} we plot the characteristic frequency   of the narrow low-frequency QPOs ($Q \gtrsim 2.5$),
 which have characteristic frequency  $\nu_{max}$ between $\nu_b$ and $\nu_h$,
 versus $\nu_h$.  
Such narrow QPOs were previously reported in other sources \citep[e.g.][and references within]{Straaten03,Straaten05} 
and we also detect them in 4U 1820--30.
Following \citet{Straaten03}, for clarity we have omitted these QPOs ($L_{LF}$) from Figure \ref{fig:nuvsnu}.
In Figure \ref{fig:bumps}, the data of 4U 1820--30 seem to follow the 
power law fitted to the $\nu_{LF}$ vs. $\nu_h$ relation of the low-luminosity
 bursters 1E 1724--3045, GS 1826--24  and the Black Hole Candidate (BHC) GX 339--4 by \citet{Straaten03,Straaten05}; therefore 
we identify these QPOs as being the $L_{LF}$ component.

\section{Discussion}\label{sec:discussion}

We have performed the first detailed study of the fast time variability in the island state of the atoll source 4U 1820--30.
It has been reported before that 
the frequencies of the variability components of the atoll sources follow a universal scheme of correlations when plotted versus
$\nu_u$ \citep[][and references within]{Straaten03}. In Figure \ref{fig:nuvsnu} (left) we show that our data are in general 
agreement with this scheme. Van Straaten et al. (2005)
 showed that the accreting millisecond pulsar SAX J1808.4--3658 
shows similar relations between its characteristic frequencies as the atoll sources do, but shifted (Figure \ref{fig:nuvsnu} - right).
This shift was interpreted to occur only between the characteristic frequencies of the low frequency components
on one hand and $\nu_u$ (and $\nu_{\ell}$) on the other, 
where $\nu_u$ (and $\nu_{\ell}$) had to be
multiplied by $\sim1.45$ to make the correlations coincide. 
Figures \ref{fig:b} and \ref{fig:h} suggest that this could also be the case for 4U 1820--30. However, 
the shift factor for $\nu_u$ is $1.21\pm0.02$ and $1.13\pm0.01$ for $L_b$ and $L_h$, respectively, 
giving an average of $1.17\pm0.01$ which is smaller than the values
of $1.420\pm0.013$ and $1.481\pm0.013$, respectively, giving an average of $1.454\pm0.009$ \citep{Straaten05}.
Similar shift factors as we find for 4U 1820--30 may in fact be present in other accreting millisecond pulsars and faint burst sources;
for example in XTE J1751--305, \citet{Straaten05} found shift factors of $1.188\pm0.045$ and $1.112\pm0.042$ for $L_b$ and $L_h$, respectively.
These results are consistent with our values, however, the results for XTE J1751--305 
have larger errors. 
It is important to note that, XTE J1751--305 has a companion of 0.013-0.35 solar mass, suggesting a heated helium dwarf
\citep{Markwardt02}. Since 4U 1820--30 also has a low-mass helium dwarf, 
the similarity in frequency shifts might be related to the chemical composition of the material in the accreting disk.
However, a simple ``frequency shift--chemical composition'' relation is not evident, since the composition of
the companion stars of SAX J1808.4--3658, 4U 0614+09, 4U 1608--52 and 4U 1728--34 are uncertain. For instance, 
SAX J1808.4--3658 might have a brown dwarf, 4U 0614+09 might have an oxygen-carbon white dwarf and both
4U 1608--52 and 4U 1728--34 might have late type main sequence companions 
\citep[but see][ respectively, for discussions]{Bildsten01,Nelemans04,Wachter02,Marti98}.           

\citet{Straaten05} suggested that the measured shift factors of $\sim1.5 =\sim 3/2$ could be related with the parametric resonance models for 
kilohertz QPOs \citep[e.g.][]{Abramowicz03}, where the 2:3 frequency resonances between general relativistic orbital/epicyclic
frequencies play a central role. The average shift factor for 4U 1820--30 is $1.17\pm0.01$, so we can reject the idea that 2:3 resonances are 
the (only) cause of the shifts.

We further attempted to test the hypothesis that a multiplicative shift of frequencies is the 
right interpretation of the difference in the frequency correlations between 
SAX J1808.4--3658 and the other atoll sources. If that hypothesis is correct, we should expect
both correlations to have the same power law index within errors, and, the 
 only significant difference between the correlations would arise from the normalization of each of the power laws.
In order to quantify the differences, we performed two different fits where simultaneously a power law is fitted to the data of 
SAX J1808.4--3658, and another power law is fitted to the data of the atoll sources 4U 1608--52, 4U 0614+09 and 4U 1728--34.
Then we compare the $\chi^2/dof$ of the fits. We only use data of the $L_b$ components since $L_h$'s behavior is more complex.
If both power law indexes and normalizations are free parameters, the best fit gives a $\chi^2/dof = 60.2/30$. 
If we force both power laws to have the same index, but different normalizations, the best fit gives a $\chi^2/dof = 86.2 / 31$.

By comparing these results using an F-test, 
we find that the improvement of the fit when leaving all the parameters free as compared to forcing equal slopes is significant at 
the $3.4\sigma$ level.
If the ``shift'' interpretation is correct, then the slopes of both correlations should be
the same and then, we should not find a significant improvement of the fit. However, 
the fact that we are dealing with $\chi^2/dof \gtrsim 2$, reduces the statistical significance of our possible interpretations.

If we perform the same analysis between 4U 1820--30 and the atoll sources, we find 
$\chi^2_1/dof$ values of  19.4/16 and $19.6 / 17$, i.e., no significant improvement of the fit.
Therefore, in both cases the data are \textit{not} inconsistent with the hypothesis that the differences between correlations
are due to only a shift in $\nu_u$ \citep{Straaten05}.

As suggested by \citet{Straaten05}, the simplest explanation for the shift between correlations, is that there is some physical difference
between sources which affects $\nu_u$. Up to now, such shifts had only been seen in accreting pulsars and only at high confidence in 
SAX J1808.4--3658, which led to the 
suggestion that the same source property that leads to strong pulsations also affects $\nu_u$ \citep{Straaten05}.
4U 1820--30 has no strong pulsations \citep{Dib04}, invalidating any strict relation between these two characteristics. 
However, as the shifts in 4U 1820--30 are smaller than in SAX J1808.4-3658 and, as accidental circumstances such as 
an unfavorable viewing geometry could suppress the pulsations in 4U 1820--30, it is too early to reject this idea. 

\textbf{Acknowledgments:}
DA likes to thank M. Linares and R. Wijnands for helpful discussions.
This work was supported by 
the ``Nederlandse Onderzoekschool Voor Astronomie'' (NOVA), i.e.,  
the ``Netherlands Research School for Astronomy'',
 and it has made use of data obtained through the High Energy Astrophysics
 Science Archive Research Center Online Service, provided by the NASA/Goddard Space Flight Center.

\begin{figure}[!hbtp] 
\center
\resizebox{0.7\columnwidth}{!}{\rotatebox{0}{\includegraphics{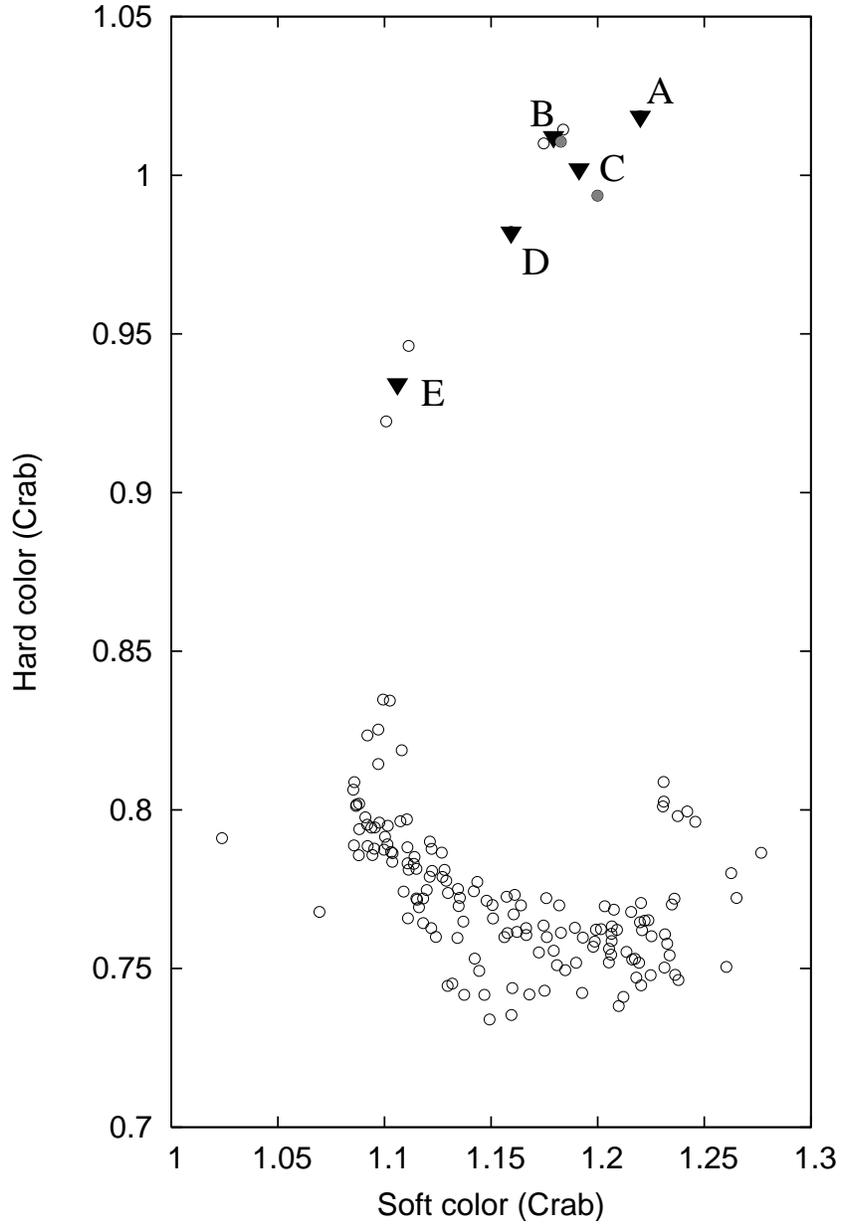}}}
\caption{4U 1820--30's hard color vs. soft color normalized to Crab colors as explained in Section \ref{sec:intro}. 
Each circle represents one of the 158 observations mentioned in Section \ref{sec:data}. 
The triangles represent the average power spectra A to E. 
They correspond to one or two observations and are labeled
in order of decreasing average hard color. For clarity, the two grey-filled circles
represent the two observations averaged to get power spectrum C.
The error bars are of the order of the size of the symbols.} 
\label{fig:ccd}
\end{figure}

\begin{figure}[!hbtp] 
\center
\resizebox{0.45\columnwidth}{!}{\rotatebox{0}{\includegraphics{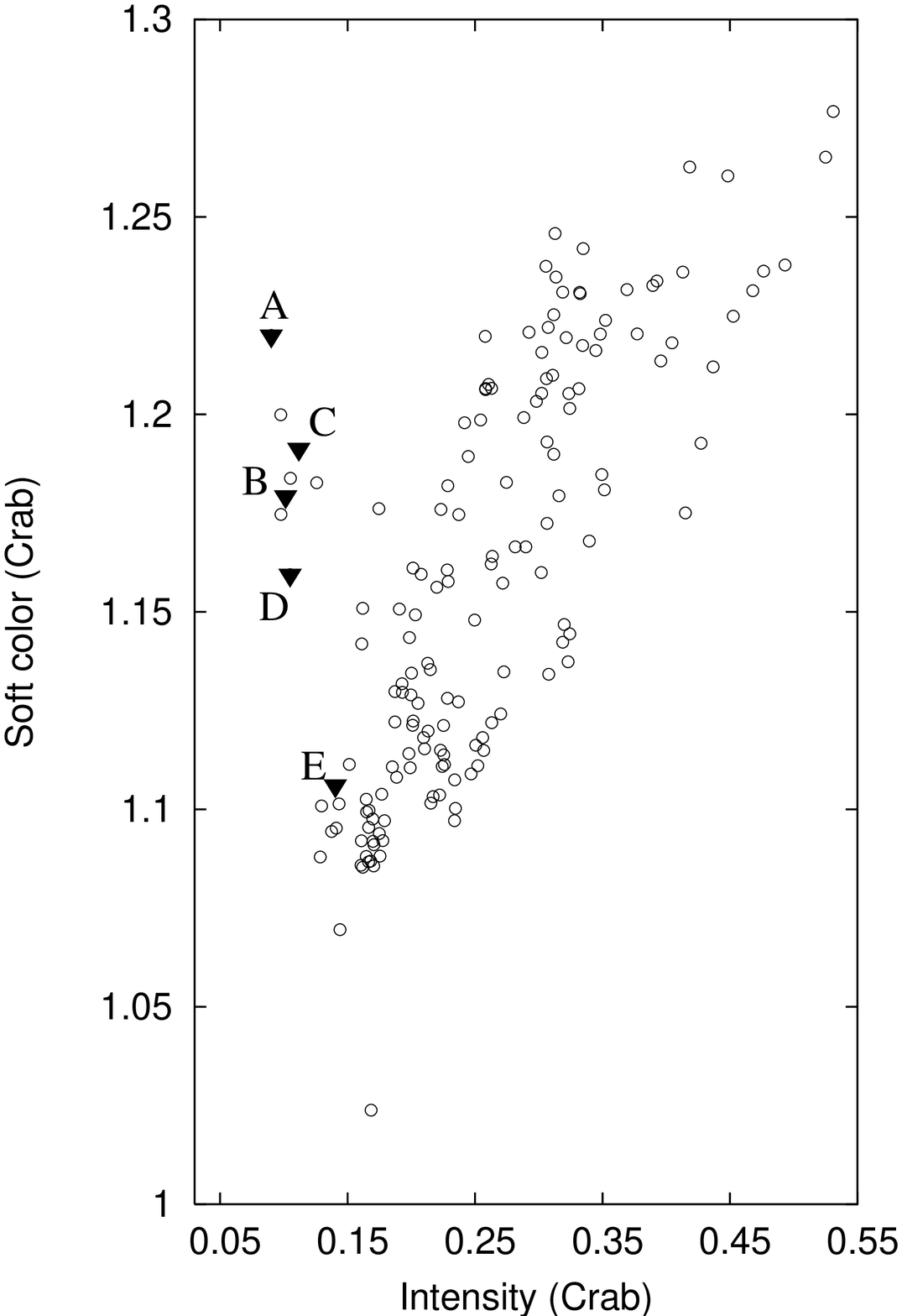}}}
\resizebox{0.45\columnwidth}{!}{\rotatebox{0}{\includegraphics{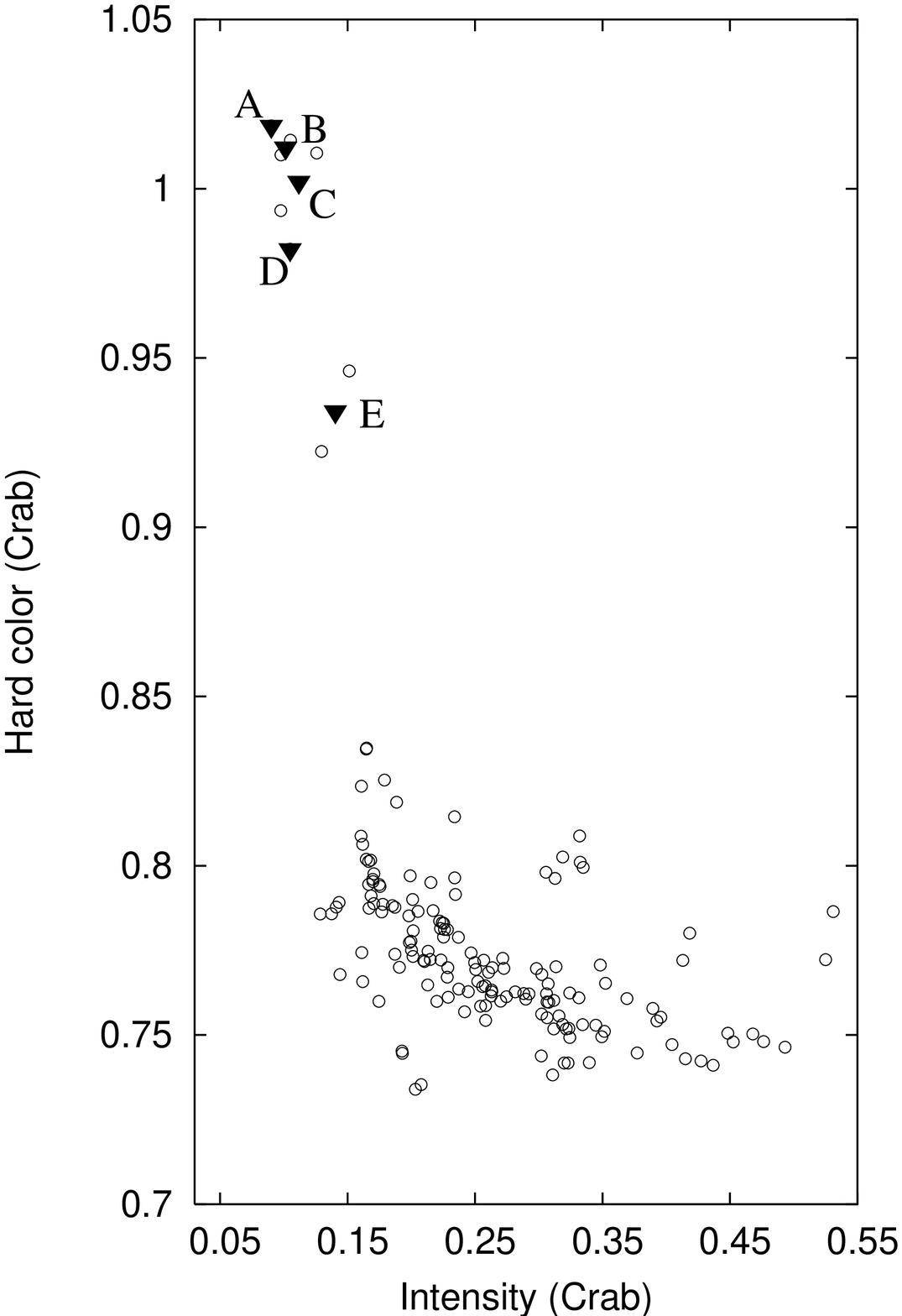}}}
\caption{Soft color vs. intensity (left) and  hard color vs. intensity (right) in Crab units as explained
in Section \ref{sec:intro}. Symbols as in Figure \ref{fig:ccd}. 
The error bars are of the order of the size of the symbols.}
\label{fig:cvsint}
\end{figure}

\begin{figure}[!hbtp]
\centering 
\resizebox{0.4\columnwidth}{!}{\rotatebox{0}{\includegraphics{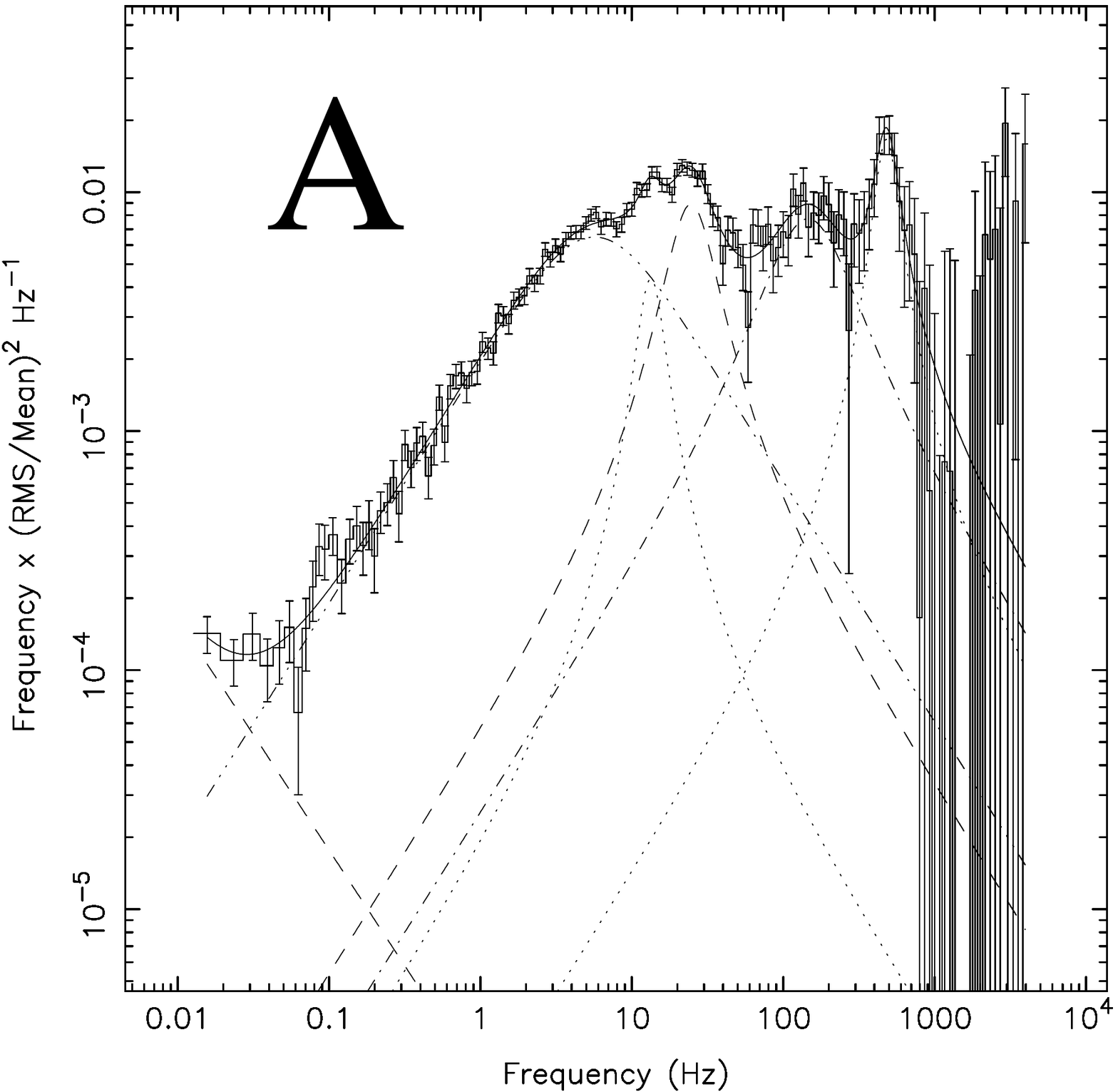}}}
\resizebox{0.4\columnwidth}{!}{\rotatebox{0}{\includegraphics{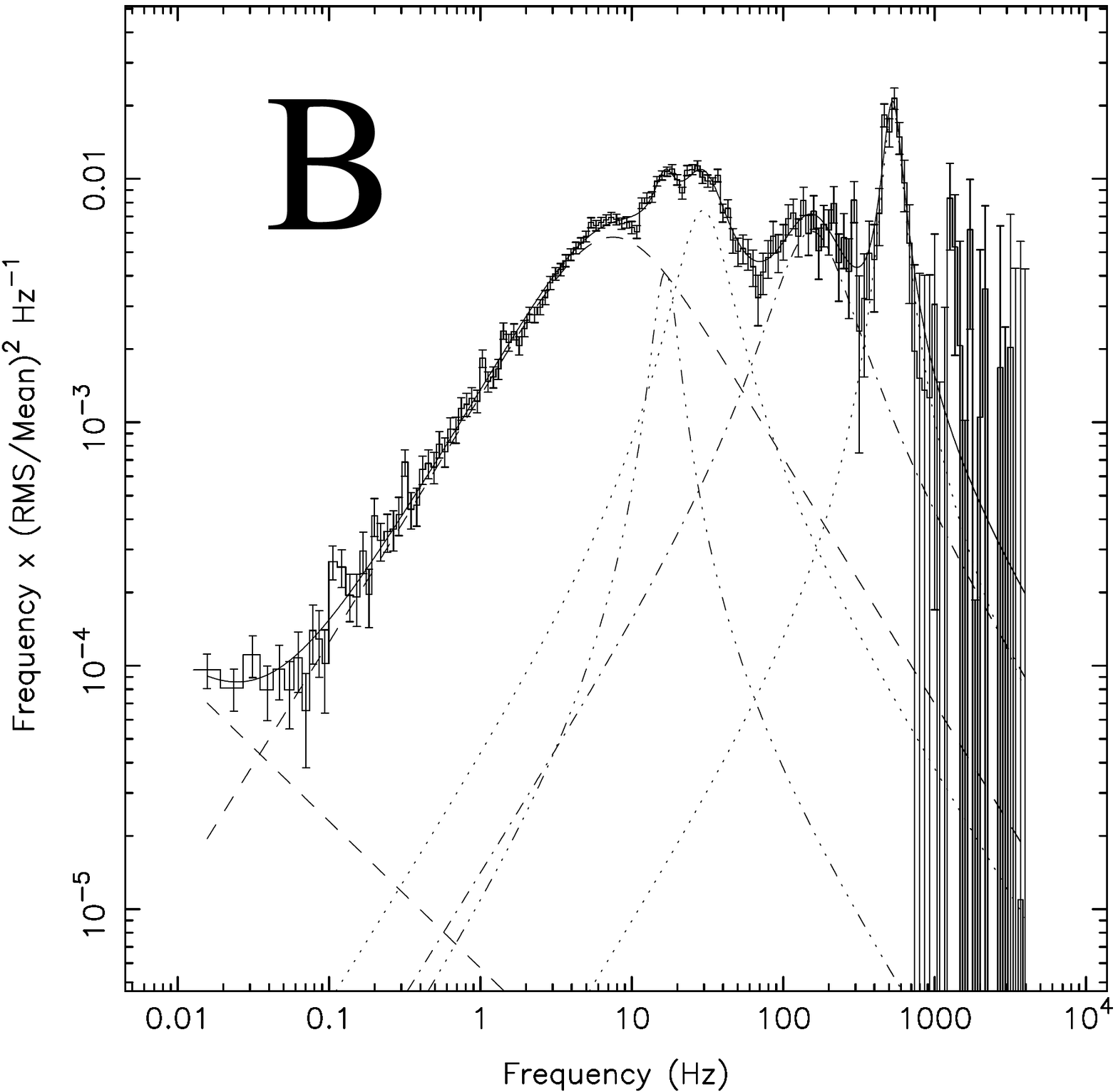}}}
\resizebox{0.4\columnwidth}{!}{\rotatebox{0}{\includegraphics{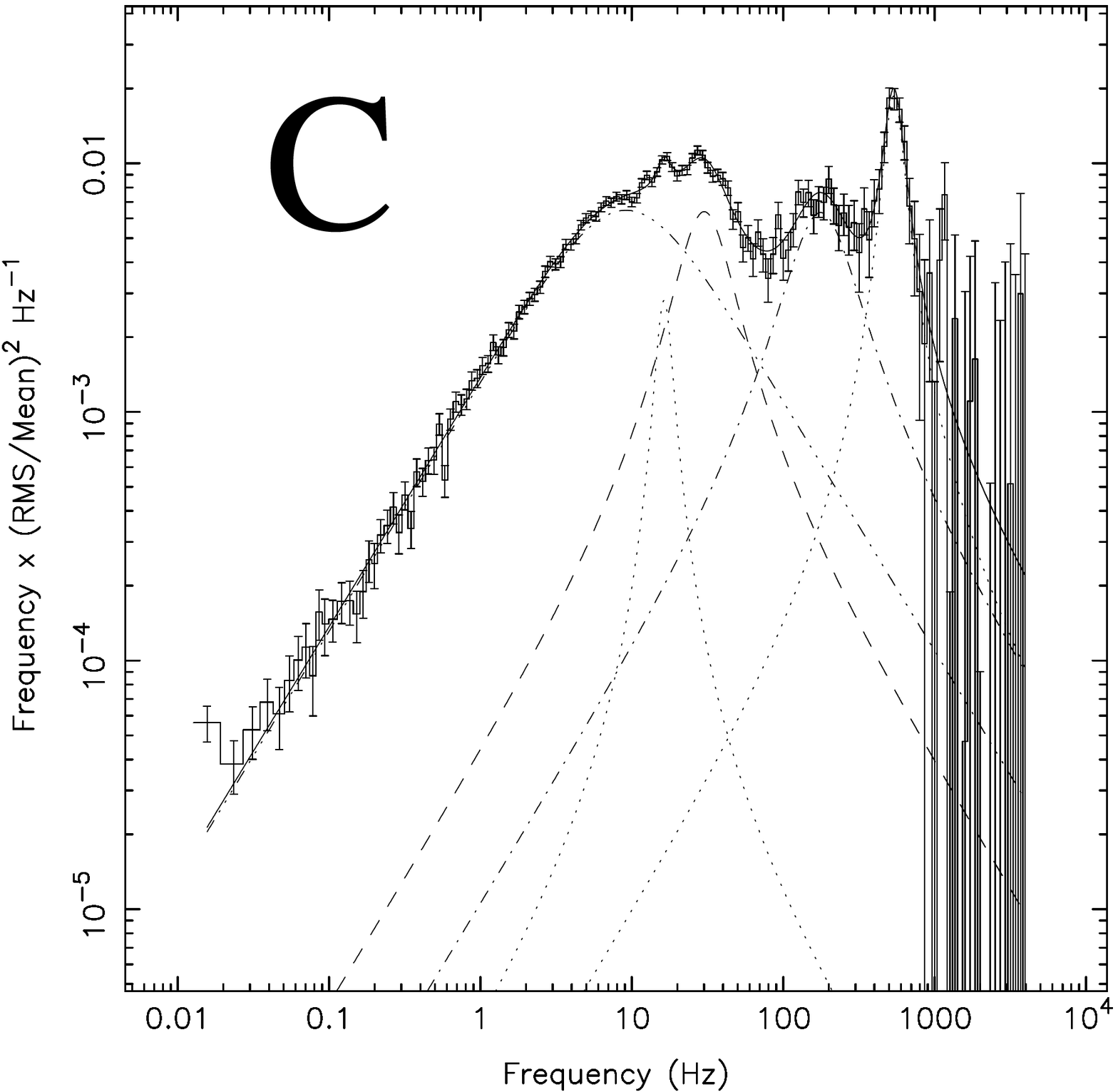}}}
\resizebox{0.4\columnwidth}{!}{\rotatebox{0}{\includegraphics{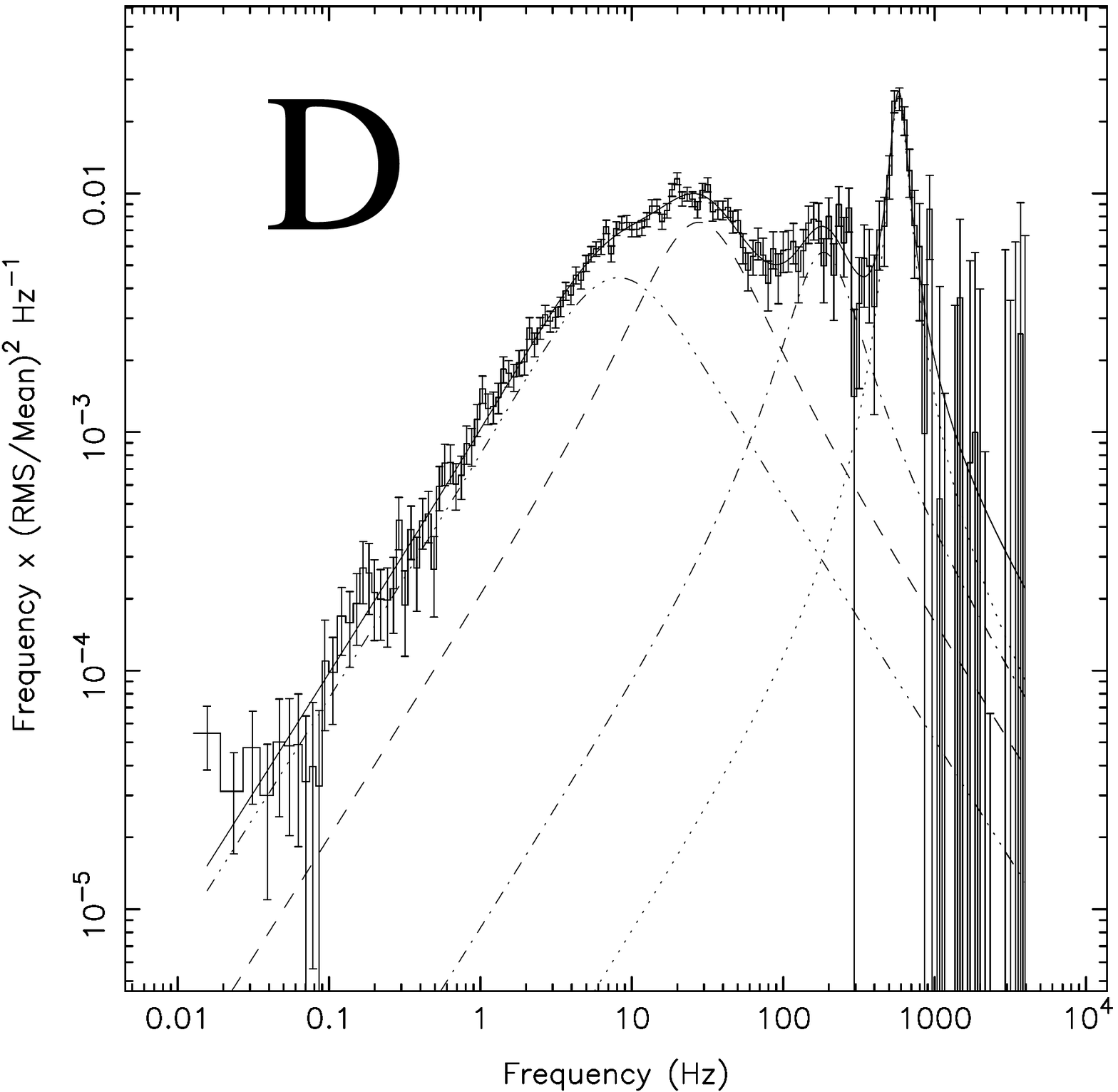}}}
\resizebox{0.4\columnwidth}{!}{\rotatebox{0}{\includegraphics{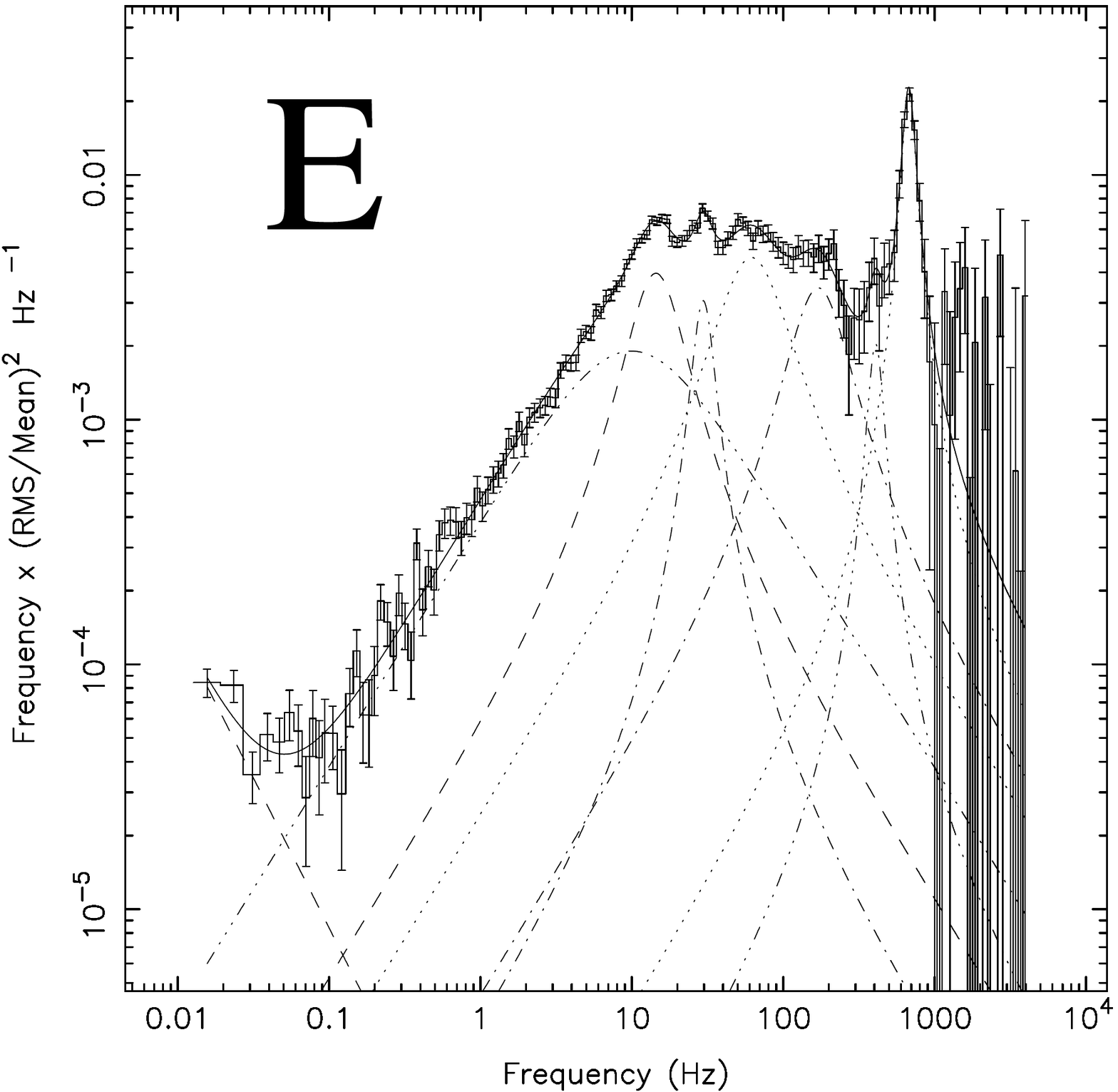}}}
\caption{Power spectra and fit functions in the power spectral density times frequency representation
for 4U 1820--30. Each plot corresponds to a different position in the color-color
and color intensity 
diagrams (see Figures \ref{fig:ccd} and \ref{fig:cvsint}).
The different lines mark the individual Lorentzian components of the fit. 
For a detailed identification, see Table \ref{table:data}, Figure \ref{fig:nuvsnu} and Section \ref{sec:results}.}
\label{fig:powerspectra}
\end{figure}

\begin{sidewaysfigure}[!hbtp] 
\center
\resizebox{0.49\columnwidth}{!}{\rotatebox{0}{\includegraphics{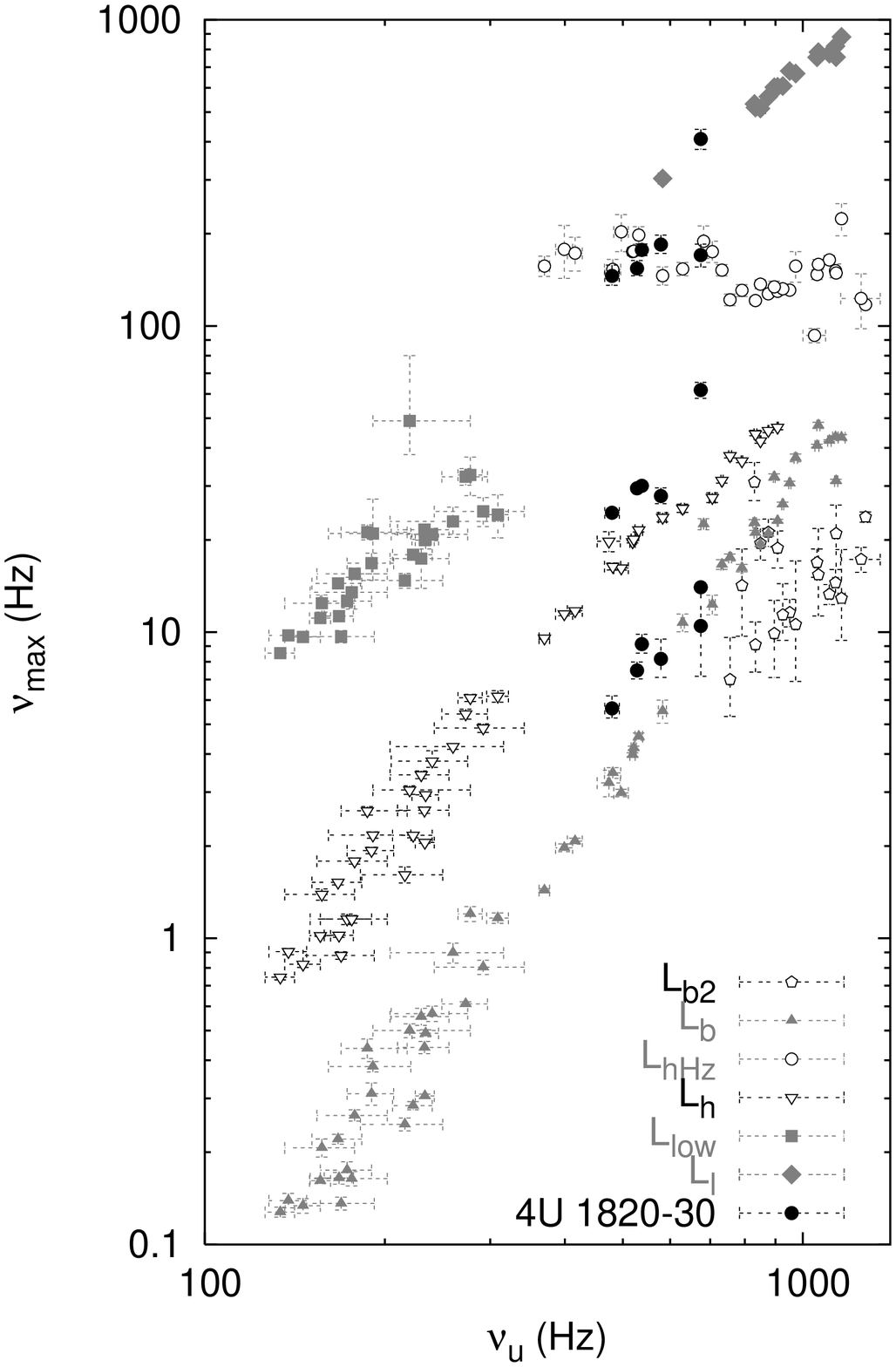}}}
\resizebox{0.49\columnwidth}{!}{\rotatebox{0}{\includegraphics{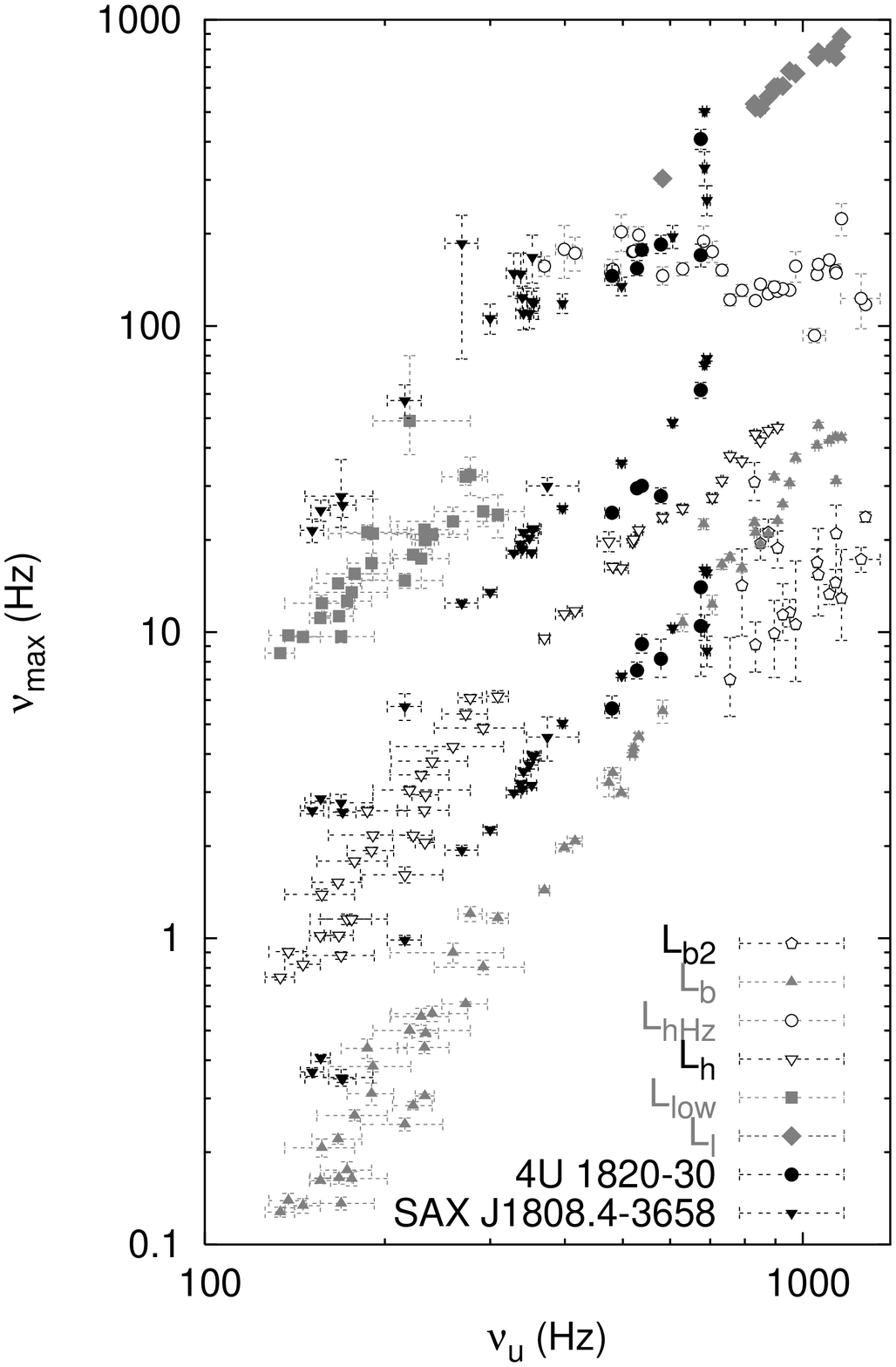}}}
\caption{Correlations between the characteristic frequencies $\nu_{max}$ of the various power
spectral components and $\nu_{u}$. For clarity, on the \textit{left} we plot the  different components 
of the atoll sources  4U 0614+09,
4U 1608--52, 4U 1728--34 and Aql X-1 and the low luminosity bursters 1E 1724--3045, GS 1826--24 and SLX
1735--269 \citep{Straaten05}, where  the black bullets mark the results for the island state features of 
4U 1820--30. On the \textit{right}, we show the same plot
as on the left, but we include the results for the millisecond accreting pulsar SAX J1808.4--3658 (black triangles).}
\label{fig:nuvsnu}
\end{sidewaysfigure}

\begin{figure}[!hbtp] 
\center
\resizebox{1\columnwidth}{!}{\rotatebox{-90}{\includegraphics{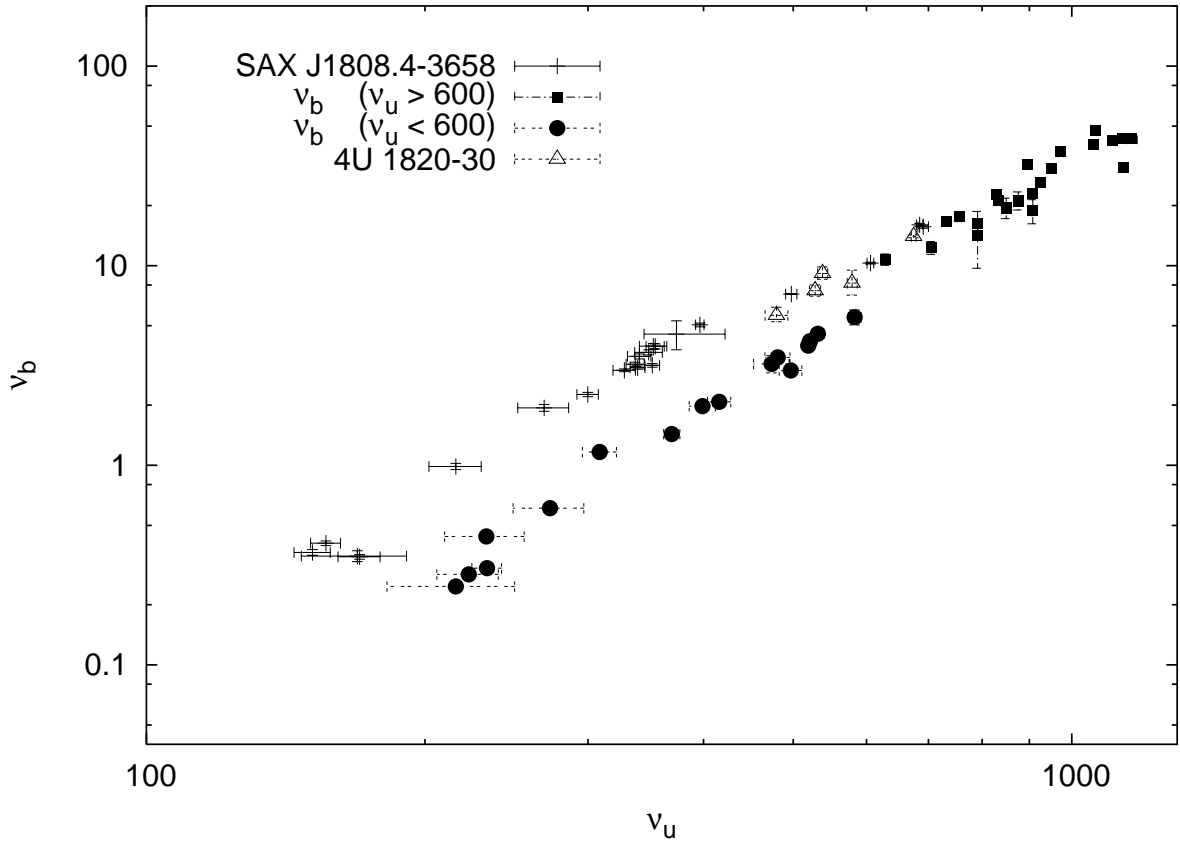}}}
\caption{Correlation between the characteristic frequencies $\nu_{b}$ and $\nu_{u}$. 
The black circles and the black squares  mark the atoll sources  4U 0614+09,
4U 1608--52, 4U 1728--34 and Aql X-1 and the low luminosity bursters 1E 1724--3045, GS 1826--24 and SLX
1735--269 \citep{Straaten05} for  $\nu_{u} < 600$ Hz  and $\nu_{u} > 600$ Hz, respectively. 
The open triangles mark the results for 4U 1820--30 and the crosses represent the results from \citet{Straaten05}
 for SAX J1808.4--3658.}
\label{fig:b}
\end{figure}

\begin{figure}[!hbtp] 
\center
\resizebox{1\columnwidth}{!}{\rotatebox{-90}{\includegraphics{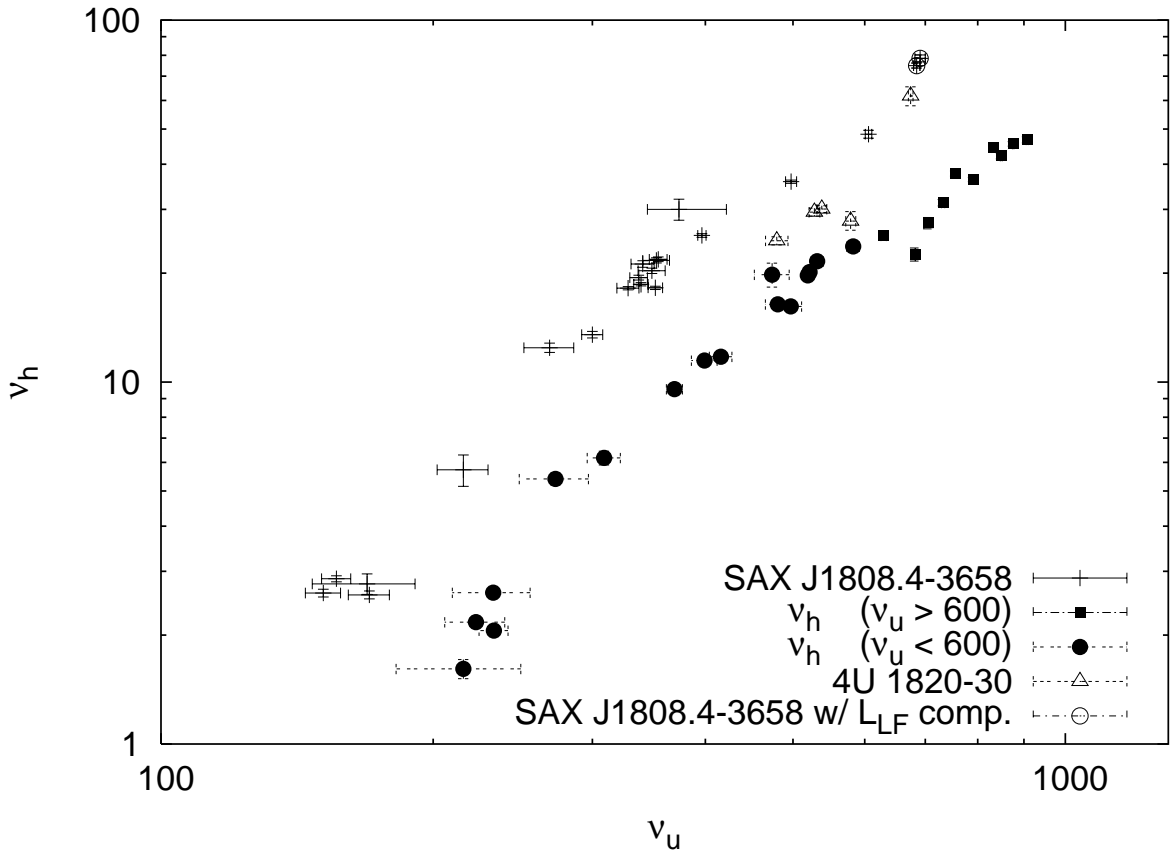}}}
\caption{Correlation between the characteristic frequencies $\nu_{h}$ and  $\nu_{u}$. 
Symbols as in Figure \ref{fig:b}. The two open circles represent the results for 
SAX J1808.4-3658 in which a $L_{LF}$ component was also found \citep[see][]{Straaten05}.}
\label{fig:h}
\end{figure}

\begin{figure}[!hbtp] 
\center
\resizebox{1.1\columnwidth}{!}{\rotatebox{-90}{\includegraphics{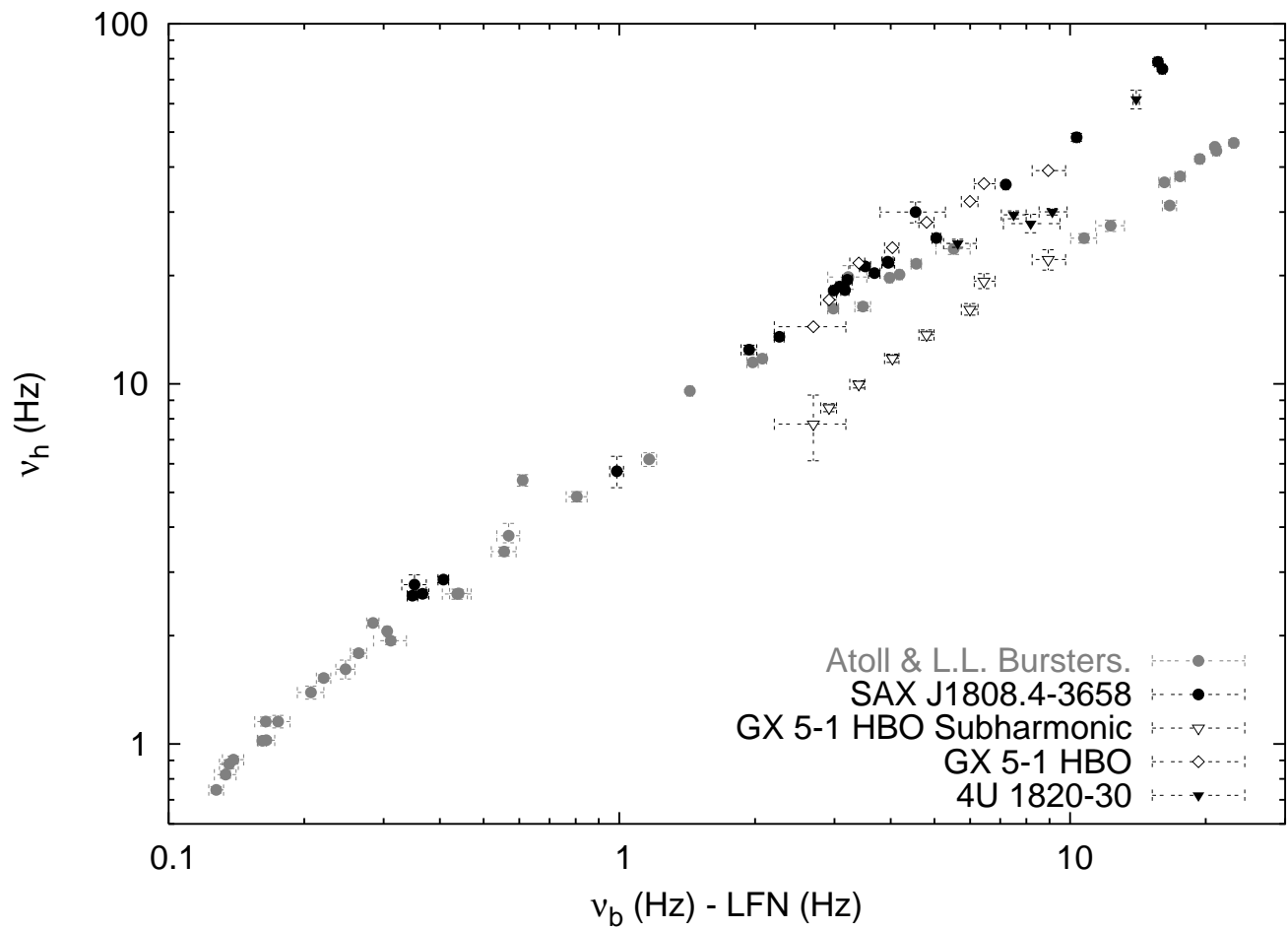}}}
\caption{The characteristic frequency $\nu_h$ plotted versus $\nu_b$. The grey circles mark the
atoll sources 4U 0614+09, 4U 1608--52, 4U 1728--34, Aql X-1 and the low luminosity bursters
1E 1724--3045, GS 1826--24 and SLX 1735--269. The black circles mark the accreting millisecond pulsar SAX 
J1808.4--3658 \citep{Straaten05}. The filled triangles mark the results for 4U 1820--30. We also include the  HBO and HBO subharmonic
characteristic frequencies of the Z-source GX 5--1 (open diamonds and open triangles, respectively),
plotted versus that of the low frequency noise (LFN) \citep{Straaten03}.
} 
\label{fig:bvsh}
\end{figure}

\begin{figure}[!hbtp] 
\center
\resizebox{1.1\columnwidth}{!}{\rotatebox{-90}{\includegraphics{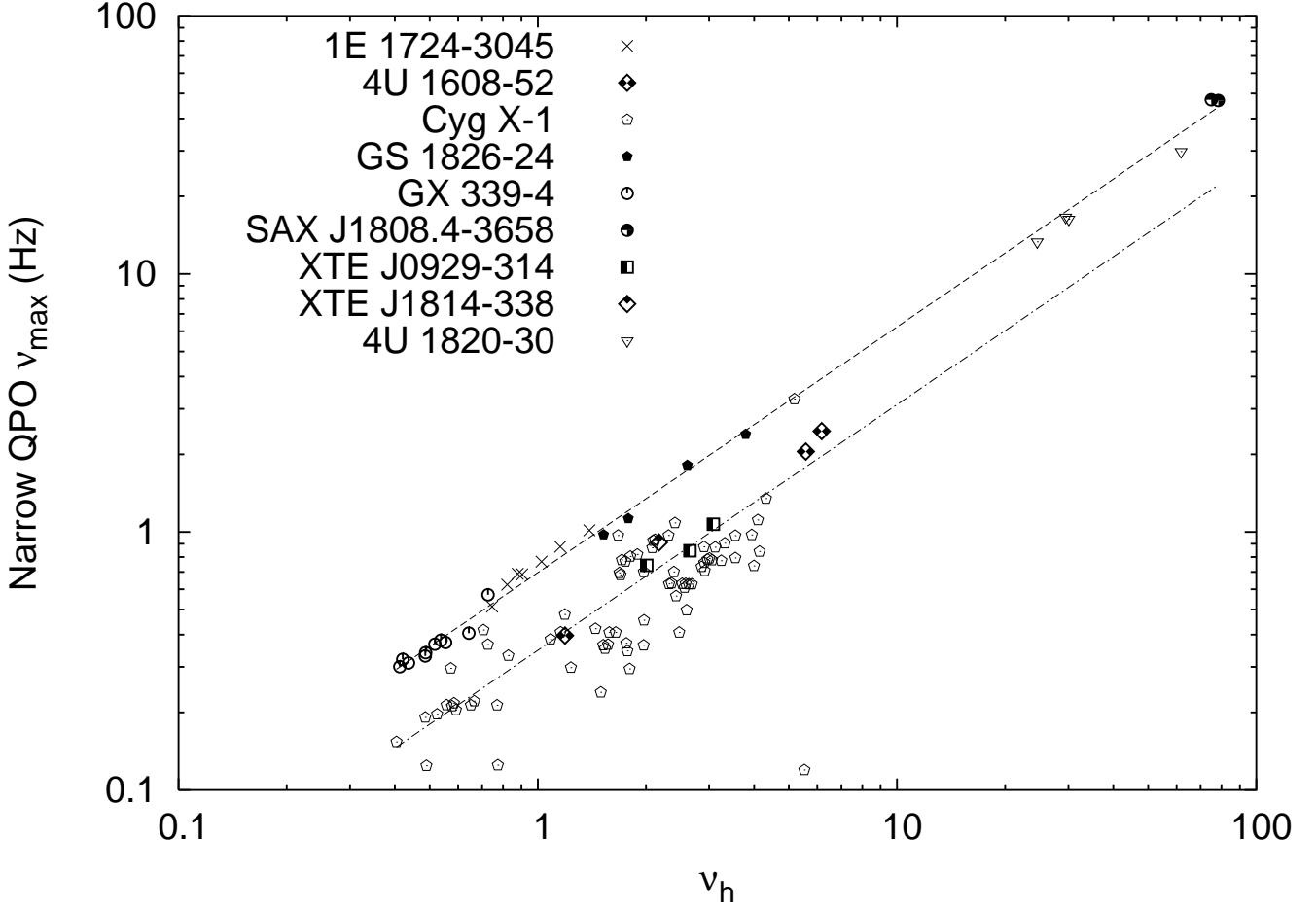}}}
\caption{Characteristic frequencies  $\nu_{LF}$ and $\nu_{LF/2}$ (see text) versus 
$\nu_h$. The symbols are labeled in the plot, and
represent the frequencies
of the QPOs from the atoll source 4U 1608--52, the BHCs Cyg X--1 and GX 339--4, the low luminosity bursters 1E 1724--3045 and GS 1826--24
 and the accreting millisecond pulsars XTE J0929--314, XTE J1814--338 and SAX J1808.4--3658 \citep{Straaten03,Straaten05}. The open
triangles show the results for 4U 1820--30.
The dashed line indicates a power law fit to the $\nu_{LF}$ vs. $\nu_h$ relation of the low-luminosity bursters 1E
1724--3045 and GS 1826--24, and the BHC GX 339--4. The dash-dotted line is a power law with a normalization half of
that of the dashed line. The error bars are of the order of the size of the symbols.}
\label{fig:bumps}
\end{figure}

{
\linespread{0.8}

\begin{table}[h]

\begin{multicols}{2}{
{\scriptsize
\begin{minipage}{1.in}
\begin{tabular}{cccc }\hline \\
   & Power spectrum A & \\
$\nu_{max}$ (Hz) & $Q$ & rms (\%)&  comp. \\
\hline \\
$ 479.73 \pm 13.65 $ & $ 2.47 \pm 0.52 $ & $ 9.90 \pm 0.82 $  & $L_u$\\
$ 145.80 \pm 11.15 $ & $ 0.58 \pm 0.17 $ & $ 11.80 \pm 0.83 $  & $L_{hHz}$\\
$ 24.56 \pm 0.63 $ & $ 1.23 \pm 0.18 $ & $ 9.80 \pm 0.67 $  & $L_h$\\
$ 13.32 \pm 0.41 $ & $ 2.15 \pm 0.56 $ & $ 5.39 \pm 0.84 $  & $L_{LF}$\\
$ 5.64 \pm 0.48 $ & $ 0.09 \pm 0.04 $ & $ 13.77 \pm 0.47 $  & $L_b$\\
\hline 
\hline \\
\end{tabular}
\end{minipage}

\begin{minipage}{1.in}
\begin{tabular}{cccc }\hline \\
   & Power spectrum B & \\
$\nu_{max}$ (Hz) & $Q$ & rms (\%)&   comp.\\
\hline \\
$ 527.99 \pm 7.32 $ & $ 3.25 \pm 0.41 $ & $ 9.46 \pm 0.45 $ & $L_u$\\
$ 154.20 \pm 8.80 $ & $ 0.72 \pm 0.14 $ & $ 9.81 \pm 0.53 $ & $L_{hHz}$\\
$ 29.48 \pm 0.74 $ & $ 1.16 \pm 0.14 $ & $ 9.15 \pm 0.52 $  & $L_h$\\
$ 16.60 \pm 0.35 $ & $ 2.50 \pm 0.47 $ & $ 4.91 \pm 0.60 $  & $L_{LF}$\\
$ 7.49 \pm 0.48 $ & $ 0.10 \pm 0.04 $ & $ 12.94 \pm 0.35 $  & $L_b$\\
\hline 
\hline \\
\end{tabular}
\end{minipage}
}
}
\end{multicols}

\begin{multicols}{2}{
{\scriptsize
\begin{minipage}{1.in}
\begin{tabular}{cccc }\hline \\
   & Power spectrum C & \\
$\nu_{max}$ (Hz) & $Q$ & rms (\%)&  comp. \\
\hline \\
$ 537.98 \pm 6.68 $ & $ 2.98 \pm 0.34 $ & $ 9.61 \pm 0.41 $  & $L_u$\\
$ 177.08 \pm 7.90 $ & $ 0.82 \pm 0.14 $ & $ 9.61 \pm 0.47 $  & $L_{hHz}$\\
$ 30.06 \pm 0.64 $ & $ 1.05 \pm 0.13 $ & $ 8.83 \pm 0.49 $  & $L_h$\\
$ 16.30 \pm 0.28 $ & $ 3.66 \pm 1.00 $ & $ 3.39 \pm 0.52 $  & $L_{LF}$\\
$ 9.14 \pm 0.66 $ & $ 0.04 \pm 0.03 $ & $ 14.05 \pm 0.40 $  & $L_b$\\
\hline 
\hline \\
\end{tabular}
\end{minipage}

\begin{minipage}{1.in}
\begin{tabular}{cccc }\hline \\
   & Power spectrum D & \\
$\nu_{max}$ (Hz) & $Q$ & rms (\%)&  comp. \\
\hline \\
$ 578.98 \pm 7.30 $ & $ 3.71 \pm 0.47 $ & $ 10.14 \pm 0.48 $  & $L_u$\\
$ 184.59 \pm 12.70 $ & $ 0.86 \pm 0.23 $ & $ 8.93 \pm 0.75 $  & $L_{hHz}$\\
$ 27.86 \pm 1.65 $ & $ 0.41 \pm 0.11 $ & $ 12.73 \pm 1.23 $  & $L_h$\\
$ 8.18 \pm 1.19 $ & $ 0.15 \pm 0.05 $ & $ 11.06 \pm 1.18 $  & $L_b$\\
\hline 
\hline \\
\end{tabular}
\end{minipage}
}
}
\end{multicols}

\begin{multicols}{2}{
{\scriptsize

\begin{minipage}{1.in}
\begin{tabular}{cccc }\hline \\
   & Power spectrum E & \\
$\nu_{max}$ (Hz) & $Q$ & rms (\%) & comp.\\ 
\hline \\
$ 675.01 \pm 4.06 $ & $ 4.58 \pm 0.41 $ & $ 8.54 \pm 0.27 $  & $L_u$\\
$ 407.94 \pm 30.54 $ & $ 3.85^{+5.13}_{-1.82}  $ & $ 2.81 \pm 0.85 $  & $L_{\ell}$\\
$ 170.37 \pm 14.67 $ & $ 0.98 \pm 0.42 $ & $ 6.65 \pm 1.44 $  & $L_{hHz}$\\
$ 61.69 \pm 3.69 $ & $ 0.77 \pm 0.26 $ & $ 8.33 \pm 1.27 $  & $L_{h}$\\
$ 29.61 \pm 0.46 $ & $ 2.80 \pm 0.53 $ & $ 4.00 \pm 0.48 $  & $L_{LF}$\\
$ 14.41 \pm 0.25 $ & $ 1.06 \pm 0.25 $ & $ 6.89 \pm 1.33 $  & $L_{b}$\\
$ 9.94 \pm 3.77 $ & $ 0.00 \pm 0.00 $ & $ 7.73 \pm 1.77 $  & $L_{b2}$\footnote{Lorentzian with $\sim 2.7\sigma$ significance }\\
\hline 
\hline \\
\end{tabular}
\end{minipage}
}

{\scriptsize
\center
\begin{minipage}{1.in}
\begin{tabular}{cccc }\hline \\
   && Power laws Parameters  \\
Power spectrum & PL index & rms (\%) & Integration \\ 
         &          &          & Interval (Hz)\\
\hline \\
A &  $1.9\pm0.5$  &  $1.19\pm0.15$ & 0.01 -- 0.08  \\ 
B &  $1.6\pm0.4$  &  $1.2\pm0.1$ & 0.01 -- 0.06  \\
E &  $2.22\pm0.24$  &  $1.02\pm0.53$ & 0.01 -- 0.08 \\
\hline 
\hline \\
\end{tabular}
\end{minipage}
}
}
\end{multicols}
\caption{Characteristic frequencies $\nu_{max}$, $Q$ values ($\equiv \nu_{0}/FWHM$ -- see Section 
\ref{sec:data}), Integrated fractional
rms (of the full PCA energy band) and identification (comp.) of the Lorentzians fitted for 4U 1820--30.
The quoted errors in   $\nu_{max}$, $Q$ and rms use $\Delta\chi^2 = 1.0$.}
\label{table:data}
\end{table}
}

\end{document}